\documentclass[10pt,twocolumn]{article}
\usepackage[letterpaper,margin=0.72in,columnsep=0.24in]{geometry}
\usepackage{amsmath,amssymb,booktabs,tabularx,array,graphicx,microtype,xcolor,enumitem,multirow,float,placeins}
\usepackage{url}
\usepackage[numbers,sort&compress]{natbib}
\usepackage[colorlinks=true,allcolors=blue!55!black]{hyperref}
\usepackage{xspace}
\usepackage{flushend}
\setlist[itemize]{label=--,nosep,leftmargin=*}
\setlist[enumerate]{nosep,leftmargin=*}
\newcolumntype{Y}{>{\raggedright\arraybackslash}X}
\newcolumntype{Z}{>{\centering\arraybackslash}X}
\definecolor{accentblue}{RGB}{20,65,130}
\definecolor{darkgray}{RGB}{60,60,60}
\newcommand{\sys}{Orvlet\xspace}

\title{\textbf{Calm Agent: Reframing Agent Interaction around Intermittent Human Attention}}

\author{%
Dian Jin\textsuperscript{1,2},
Zixun Fu\textsuperscript{3},
Guan Wang\textsuperscript{3},
Yancheng Yuan\textsuperscript{3}\thanks{Corresponding author: Yancheng Yuan.}
\\[0.4em]
\small\textsuperscript{1}Research Institute for Intelligent Wearable Systems,
The Hong Kong Polytechnic University\\
\small\textsuperscript{2}School of Fashion and Textiles,
The Hong Kong Polytechnic University\\
\small\textsuperscript{3}Department of Applied Mathematics,
The Hong Kong Polytechnic University
}
\date{}

\makeatletter
\renewcommand\paragraph{\@startsection{paragraph}{4}{\z@}%
  {3.25ex \@plus 1ex \@minus .2ex}{-1em}{\normalfont\normalsize}}
\makeatother

\begin{document}
\maketitle

\begin{abstract}
AI agents can carry out auxiliary tasks while users focus on other work. Yet users still need to understand task state, make changes, and respond when help is required. We argue that agents for ongoing auxiliary tasks should be designed around intermittent human attention: tasks should continue between interactions, and users should be able to return with little effort to understand and control them. We develop this position as Calm Agent. It connects three on-demand interaction modes: Delegate, Adjust, and Check, through a persistent task representation. Generative interaction provides controls and feedback suited to open-ended goals, while existing controls support routine actions. We illustrate these principles with Orvlet, a desktop prototype for background execution and compact task interfaces. We identify challenges in adapting interfaces while preserving continuity, coordinating attention across tasks, providing responsive interaction, and learning user habits over time. The aim is to support continuing task control while protecting attention for primary work.
\end{abstract}

\section{Introduction}
\label{sec:intro}

AI agents currently can use tools and complete complex multi-step tasks in browsers and desktop applications \cite{yao2022react,zheng2024gpt}. These capabilities allow users to delegate auxiliary tasks while continuing their primary work. For example, a user writing a paper may also want background music or a collaborator's local time. Such tasks can remain active for a long time, although the user needs to interact with each one only briefly.

\begin{figure}[t]
    \centering
    \includegraphics[width=\columnwidth]{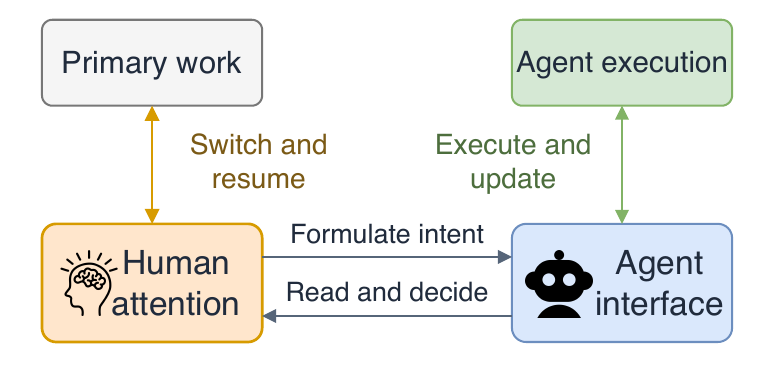}
    \caption{The human attention bottleneck. Primary work and agent interaction require the same limited and valuable attention. }
    \label{fig:bottleneck}
\end{figure}

Delegating execution leaves an important interaction problem. Users  need to open a new agent interface, read earlier messages to understand the current state, and describe the need for change in natural language. These actions actually share the  limited and valuable attention on primary work (Fig.~\ref{fig:bottleneck}). Research on interruptions \cite{mark2008cost,iqbal2007disruption} also shows that switching tasks can add extra effort and require recovery when attention return. As agents take on more ongoing tasks, the way users access and manage them becomes an important design concern.

Our position is that a agent for ongoing auxiliary tasks should be designed to avoid capturing too much valuable human attention. A task should remain available and continue within its delegated scope while the user focus on their promary work. When the user returns to the auxiliary task, the current state of agent and its relevant controls should be directly accessible rather than reconstructing the task from a conversation. When a user input is necessary, the agent should make this need apparent. 
This position focus on how tasks continue between interactions and how easily users can return to understand and control them.

Calm Technology \cite{weiser1996designing,case2015calm} provides a concept foundation: technology should move between the center and periphery of attention as needed. Open-ended agent delegation adds a further requirement: The system need to interpret a goal, identify suitable actions to achieve the goal, receive feedback from the environment, and keep these parts connected to execution as the task develops. Generative UI \cite{leviathan2026generative,chen2026generative,cao2025generative} can help provide task-specific interfaces, while language and direct manipulation offer complementary ways to express goals and make precise changes \cite{masson2024directgpt,subramonyam2024bridging}. These capabilities need to support a coherent relationship with a continuing task.

We develop this position as Calm Agent. We define three interaction modes for Calm Agent: Delegate, Adjust, and Check. These modes give users access to the same task through language, direct controls, and state feedback. A shared task representation connects these interactions to background execution. The framework identifies which task information and controls should remain available while users focus on other work. 
We illustrate the framework through \sys\footnote{\url{https://github.com/CLaSLoVe/Orvlet}}, a desktop prototype, and use its design to develop questions for further research. This paper contributes:
\begin{enumerate}
    \item A position and design framework for agent interaction under intermittent human attention, connecting continuous task execution with users’ ability to inspect and control tasks on demand.
    \item A prototype Orvlet shows how background execution, generated task interfaces, and reusable controls can support this framework.
    \item An analysis of open challenges in interface adaptation and continuity, attention requests across concurrent tasks, responsive generation and execution, and long-term user adaptation.
\end{enumerate}

\section{Related Work}
\label{sec:related}

\subsection{Calm Technology and Attention}
Weiser and Brown's Calm Technology \cite{weiser1996designing} describes movement between the center and periphery of attention. Peripheral Interaction \cite{bakker2015peripheral} and Calm Design Patterns \cite{case2015calm} explore how information and simple actions can remain accessible with limited attention. Recent work on ambient analytics connects peripheral awareness with opportunities for closer inspection \cite{hubenschmid2026ambient}.

The SEEV model \cite{steelman2017theory} explains visual attention allocation through Salience, Effort, Expectancy, and Value. It connects the visual prominence of information and the effort of attending to it with the expected occurrence of new information and its relevance to the task. These factors provide a useful perspective on access to background tasks.

These previous research explains how interaction can fit around a primary activity. However, open-ended delegation adds a problem of determining what the peripheral interface should contain. The relevant state and actions depend on the delegated goal and can change during execution. Therefore, an agent interface must connect peripheral presentation to   goals and  available actions. Calm Agent addresses this connection by organizing peripheral access around a persistent task that can be revised, controlled, and checked.

\subsection{Generative and Hybrid Interfaces}
Generative UI \cite{leviathan2026generative,chen2026generative} uses language models to produce interfaces for user requests. DynaVis \cite{vaithilingam2024dynavis} generates smaller widgets for visualization editing. Jelly \cite{cao2025generative} uses an evolving task-driven data model to generate interfaces and translate natural language and direct manipulation into model changes. Its model-based approach demonstrates how interface adaptation can be organized around a persistent structure.

Direct manipulation and mixed-initiative interaction provide broader foundations for combining graphical actions with language \cite{shneiderman1983direct,horvitz1999principles}. Studies \cite{subramonyam2024bridging} identify difficulties in expressing intentions through prompts, and DirectGPT \cite{masson2024directgpt} shows how direct manipulation can reduce interaction time and prompt-writing effort.

These approaches establish how users can create and change task interfaces. For an ongoing auxiliary task, the interface must also support brief visits while execution continues elsewhere. A generated interface may fit the goal but still require users to reopen a workspace, locate changed controls, and wait for regeneration, which discouraging  its practical use. The design gap concerns the continuity of access across these visits: which controls should persist, which changes require generation, and how displayed state stays connected to execution. Calm Agent uses Delegate, Adjust, and Check to organize this relationship around the same ongoing task.

\subsection{Agent Feedback and Control}
GUI agents raise questions about how people maintain awareness and control during automated execution \cite{parasuraman2000model}. AgentLens \cite{kim2026agentlens} combines background mobile execution with selective overlays and adapts between full, partial, and generated interfaces when interaction is needed. It provides an important example of minimally invasive, just-in-time communication. Choosing a suitable presentation at an intervention point, however, leaves a further design question: how should users access the same task between interventions, including when they initiate a check or adjustment themselves? Calm Agent makes this continuing access explicit through a task representation shared by goal revision, direct control, and state inspection.

Research on proactive programming assistance examines the trade-off between useful support and interruption \cite{pu2025assistance}. Work on overhearing agents studies how agents obtain context and decide when to intervene \cite{zhu2025overhearing}. For explicitly delegated auxiliary tasks, intervention must also remain connected to the user's existing goal. A clarification, a request for a decision, and an unsolicited suggestion place different demands on attention. Calm Agent connects necessary attention requests to the same task that users can revisit on their own initiative.

Together, these research areas motivate a framework that connects open-ended delegation, persistent controls, and selective attention requests across the lifetime of a task. Calm Agent develops this framework for periods when the user is primarily occupied elsewhere. Its focus is the relationship between continuing execution and intermittent participation, including both user-initiated visits and agent-initiated requests.

\section{The Calm Agent Position}
\label{sec:paradigm}

A task may run for minutes or hours while the user gives it attention for only a few seconds at a time. We argue that agent interfaces should support this difference in duration. Users need both the ability to leave a task and an effective way to return. We develop three design principles from this position, then describe the interaction modes and task representation that support them.

\subsection{Design Principles}

\paragraph{P1: Maintain the Task between Interactions.}
Task execution and context should persist independently of whether the interface is open. Within the delegated scope, the agent should maintain the goal, update observable state, and record changes relevant to later interaction. If progress requires user input, it should preserve the task and expose that need. The user should not have to keep a conversation active to keep the task available.

\paragraph{P2: Support Brief, Purposeful Interaction.}
The interaction method should fit what the user wants to do. Language supports open-ended goals; an available control supports a specific adjustment; a state display supports inspection. Generative interaction should select useful information and actions for the task. Once a control is available, routine use should retain its meaning and avoid unnecessary interface regeneration.

\paragraph{P3: Make Returning Informative and Actionable.}
When users return, they should be able to understand the current situation and identify what they can do next. The interface should show observed state, available actions, and any need for input. Its prominence should fit the need for attention. A compact display can support this principle, but its value depends on whether users can understand and act on the task with little effort.

Together, these principles set concrete criteria for assessing a design: a task should remain available while the user is away, its state should be legible without reviewing the full history, routine adjustments should be supported directly, and necessary interventions should be noticed in time. Section~\ref{sec:challenges} examines the open challenges in meeting these requirements.

\subsection{Three On-Demand Interaction Modes}
\begin{figure}[t]
    \centering
    \includegraphics[width=0.98\columnwidth]{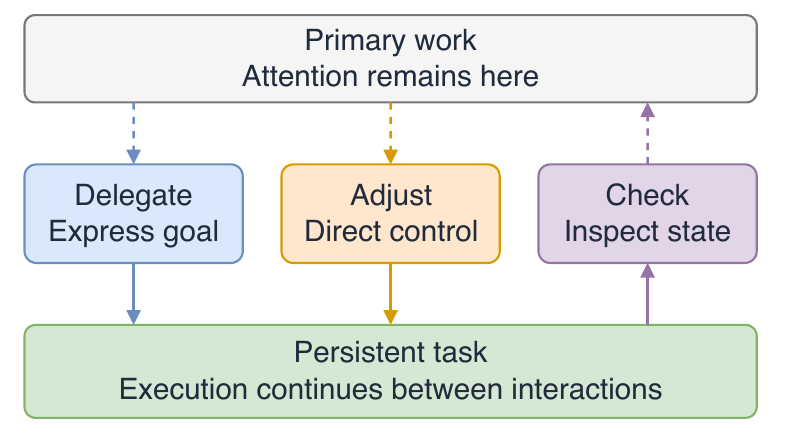}
    \caption{On-demand interaction with an ongoing task. Delegate expresses goals, Adjust changes task parameters, and Check shows task state. }
    \label{fig:timeline}
\end{figure}

Delegate, Adjust, and Check describe the user's purpose when interacting with an ongoing task (Fig.~\ref{fig:timeline}). Initial delegation establishes the task. Later interactions can use any mode, in any order, depending on the user's needs.

\paragraph{Delegate: Express or Revise a Goal.}
The user uses language to express an intention that the available controls do not capture. The agent interprets the goal or revised constraints and determines the required actions. A revision can continue an existing task while changing what it should achieve.

\paragraph{Adjust: Use an Available Control.}
The user invokes an action or changes a parameter through direct manipulation. The control has a meaning defined by the current task, such as pausing playback or changing a timer value. Adjust refers to a user action; automatic retries and runtime updates belong to background execution.

\paragraph{Check: Inspect the Current State.}
The user reads task feedback without issuing a new instruction. A brief view may be enough to establish that the task is running, waiting for input, or complete. Further detail should be available when the user needs to understand a change or decide how to respond.

\paragraph{Agent-Initiated Attention and Clarification.}
The three modes describe what the user does; an attention request describes how an interaction begins. The agent may request participation when essential information is missing or a task needs a decision. The user can then inspect the request through Check, clarify the goal through Delegate, or respond with an available control through Adjust. Agent-initiated attention therefore connects to all three modes rather than forming a fourth mode. A clarification should ask for the information needed to proceed and remain attached to the existing task. If a decision can wait, the task should retain that request until the user returns. The timing and prominence of these requests are design choices discussed in Section~\ref{sec:attention}.

\subsection{Connecting Interaction to Execution}
\begin{figure*}[t]
    \centering
    \includegraphics[width=0.9\textwidth]{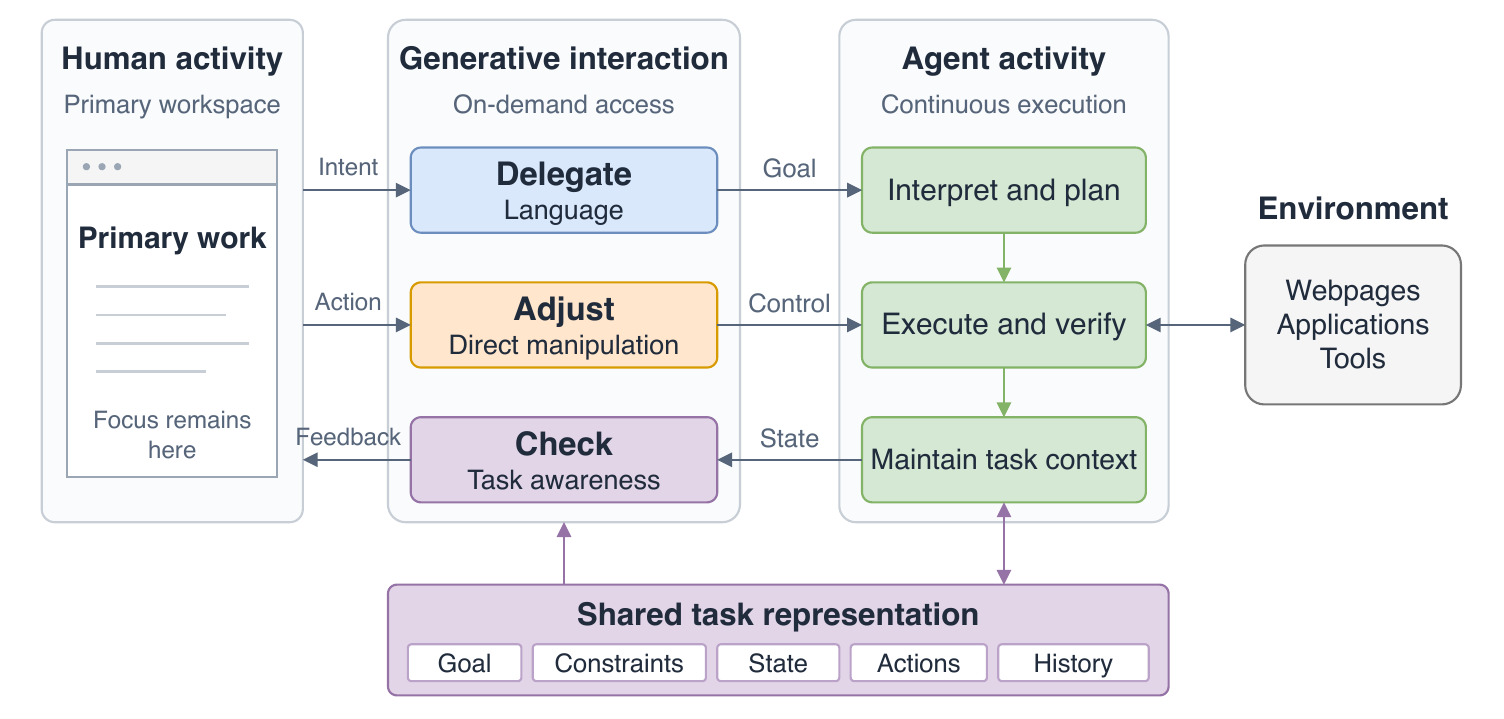}
    \caption{Conceptual architecture of Calm Agent. The shared task representation keeps the interface and execution connected while users focus on primary work.}
    \label{fig:conceptual_arch}
\end{figure*}

Fig.~\ref{fig:conceptual_arch} connects the interaction modes to the task environment. The user remains in the primary workspace. The generative interaction layer selects controls and feedback for the interpreted goal, while the agent acts through webpages, local applications, or system tools. A persistent task representation connects these activities.

We describe each task as
\begin{equation}
    T_i = \langle \mathcal{G}, \mathcal{C}, \mathcal{S}, \mathcal{A}, \mathcal{H} \rangle,
\end{equation}
where $\mathcal{G}$ is the goal, $\mathcal{C}$ contains user and environment constraints, $\mathcal{S}$ stores observable state, $\mathcal{A}$ contains available actions and their handlers, and $\mathcal{H}$ records execution history. Delegate can revise the goal and constraints; Adjust invokes an action; Check reads state. Execution can update state and history without a new user request.

This representation distinguishes changes that need interface adaptation from those that can use existing controls. When a revised goal introduces a new interaction need, generative interaction can provide suitable controls and feedback. When the user invokes an available action, the system can reuse its binding. A runtime update may only require changing a displayed value. The design principle is to adapt the interface when interaction needs change, while keeping familiar controls stable during routine use.

For example, switching background music to instrumental jazz changes the selection criteria, but does not necessarily change the need for playback controls. The agent can interpret the revised preference while retaining the meaning and location of those controls. A track change may then update the displayed title without changing the layout. For intermittent interaction, this continuity matters: users should be able to reuse what they learned during an earlier visit. If an action is no longer available, the interface should indicate that change rather than silently assign a different meaning to the same control.

The connection must also remain meaningful as the environment changes. A displayed control should refer to an action that is currently available, and feedback should reflect observed execution. If the agent cannot establish the current state, the interface should make that uncertainty visible. These requirements support the user's ability to return and act on the task.

\section{The Orvlet Prototype}
\label{sec:system}

\sys provides a concrete example of the Calm Agent principles on the desktop. It combines an isolated background browser, compact task interfaces, and stored task context. We use its design and several implemented scenarios to show how the principles shape interaction.

Fig.~\ref{fig:orvlet-overleaf} shows how a writer can access
Orvlet's input bar and music controls while editing a manuscript.

\begin{figure*}[t]
    \centering
    \includegraphics[width=0.9\textwidth]{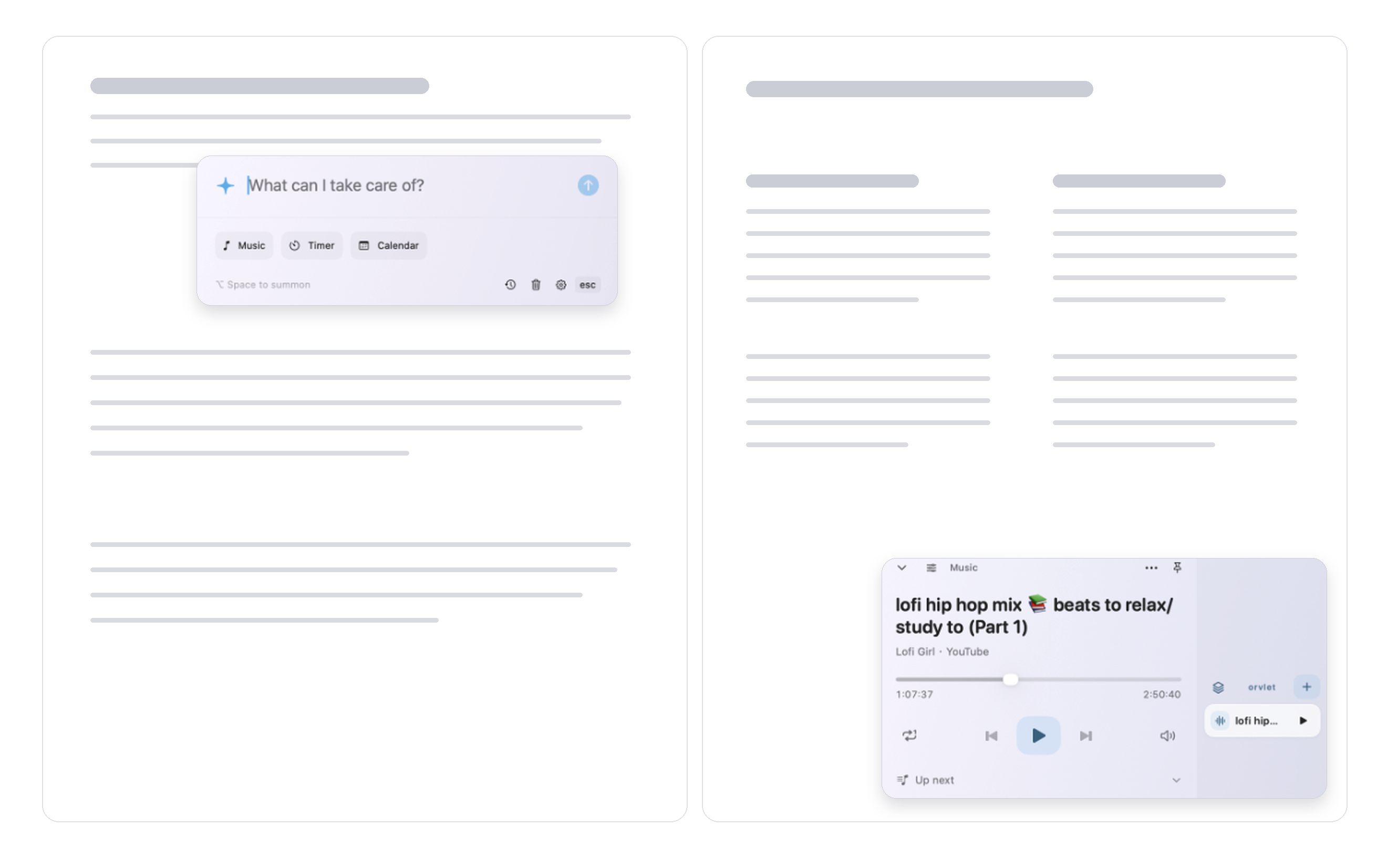}
    \caption{Orvlet during manuscript editing. The floating
input bar lets the writer delegate a task, and the expanded music
card at the lower right provides playback controls. When collapsed,
the card becomes a small pill at the edge of the workspace. After
three seconds of inactivity, the pill fades to a semi-transparent
state while the music task continues in the background.}
    \label{fig:orvlet-overleaf}
\end{figure*}

\subsection{Background Task Execution}
\sys operates the relevant webpage or tool in the background. For a media request, it performs the webpage actions and checks playback before presenting the result. The original service provides the task capability, while the compact interface provides access to selected controls and state. Collapsing this interface leaves the background activity running, supporting P1.

\begin{figure*}[t]
    \centering
    \includegraphics[width=0.9\textwidth]{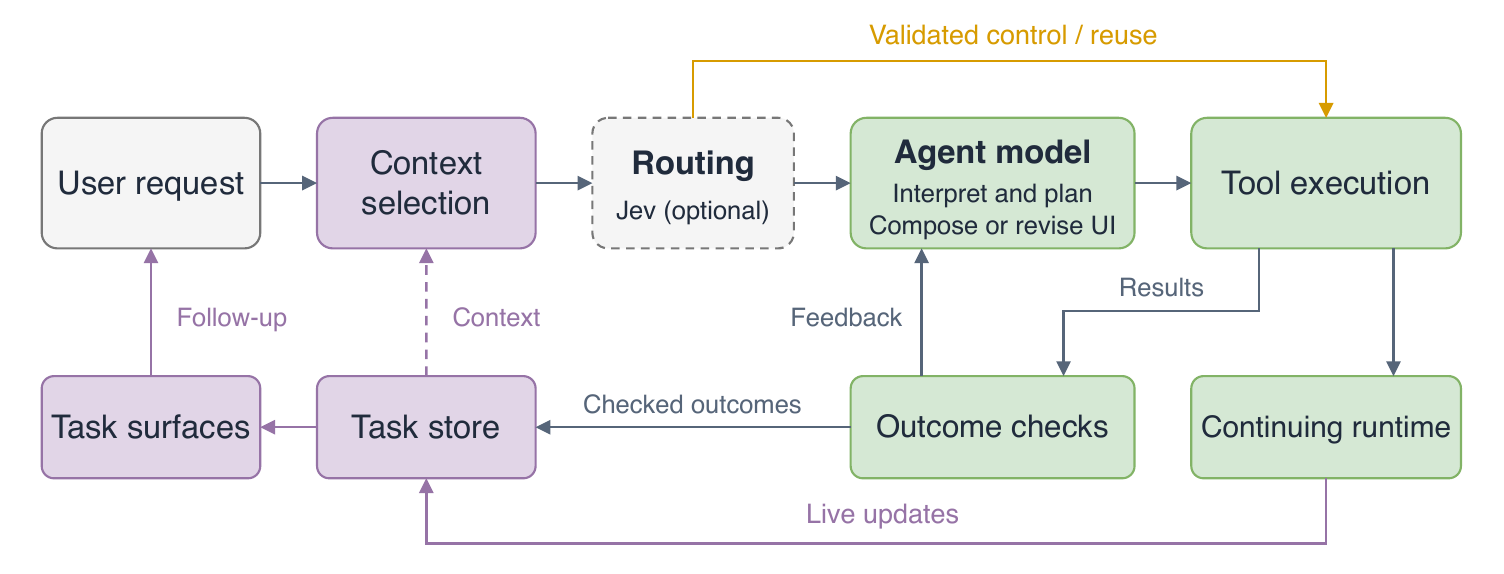}
    \caption{Task execution and interaction in \sys. Requests use the current task context, and validated controls can be reused directly. Outcome checks provide feedback and update task state. The continuing runtime also provides live updates. }
    \label{fig:sys_arch}
\end{figure*}

Fig.~\ref{fig:sys_arch} shows the execution path. New goals can require planning; existing controls invoke their action bindings. Outcome checks and runtime updates feed into task context, which supports later requests and the visible interface. The implementation can optionally use Jev \cite{duan2026general} to route requests to available capabilities  fast.

\subsection{Brief Access and Task Return}
\begin{figure*}[t]
    \centering
    \begin{tabularx}{\textwidth}{@{} Z Z @{}}
        \includegraphics[width=0.95\linewidth,height=3.4cm,keepaspectratio]{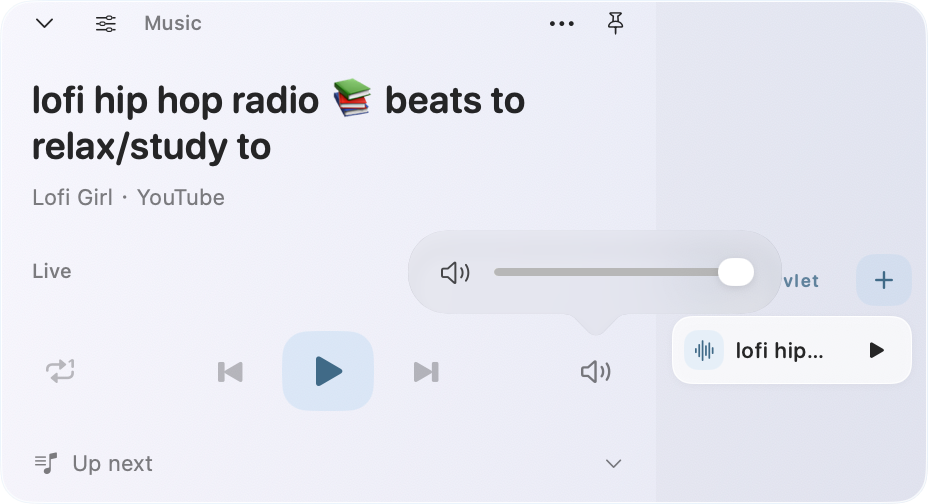} &
        \includegraphics[width=0.67\linewidth,height=2.2cm,keepaspectratio]{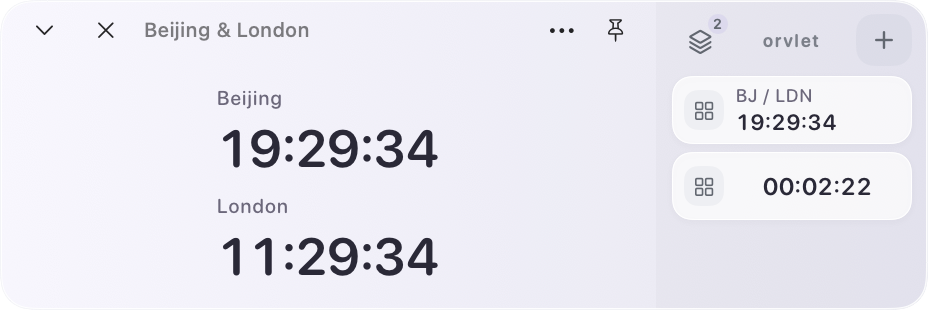} \\
        \small (a) Music Playback Controls &
        \small (b) World Clock and Countdown \\[6pt]

        \includegraphics[width=0.73\linewidth,height=3.5cm,keepaspectratio]{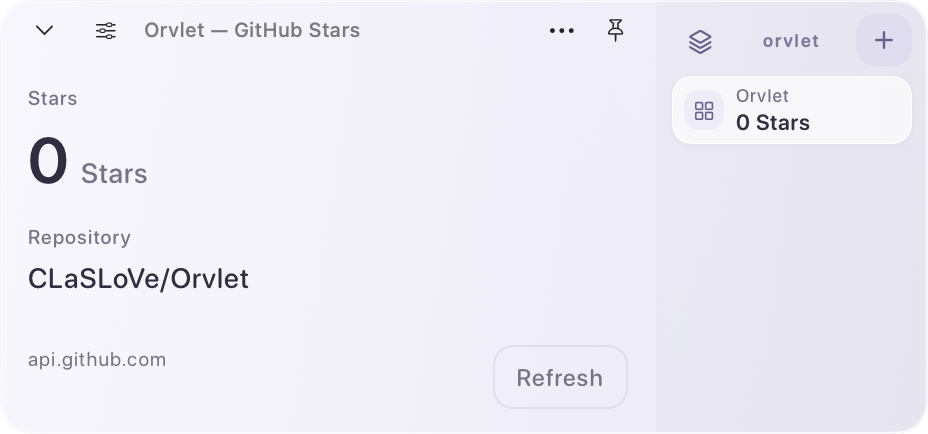} &
        \includegraphics[width=0.99\linewidth,height=3.9cm,keepaspectratio]{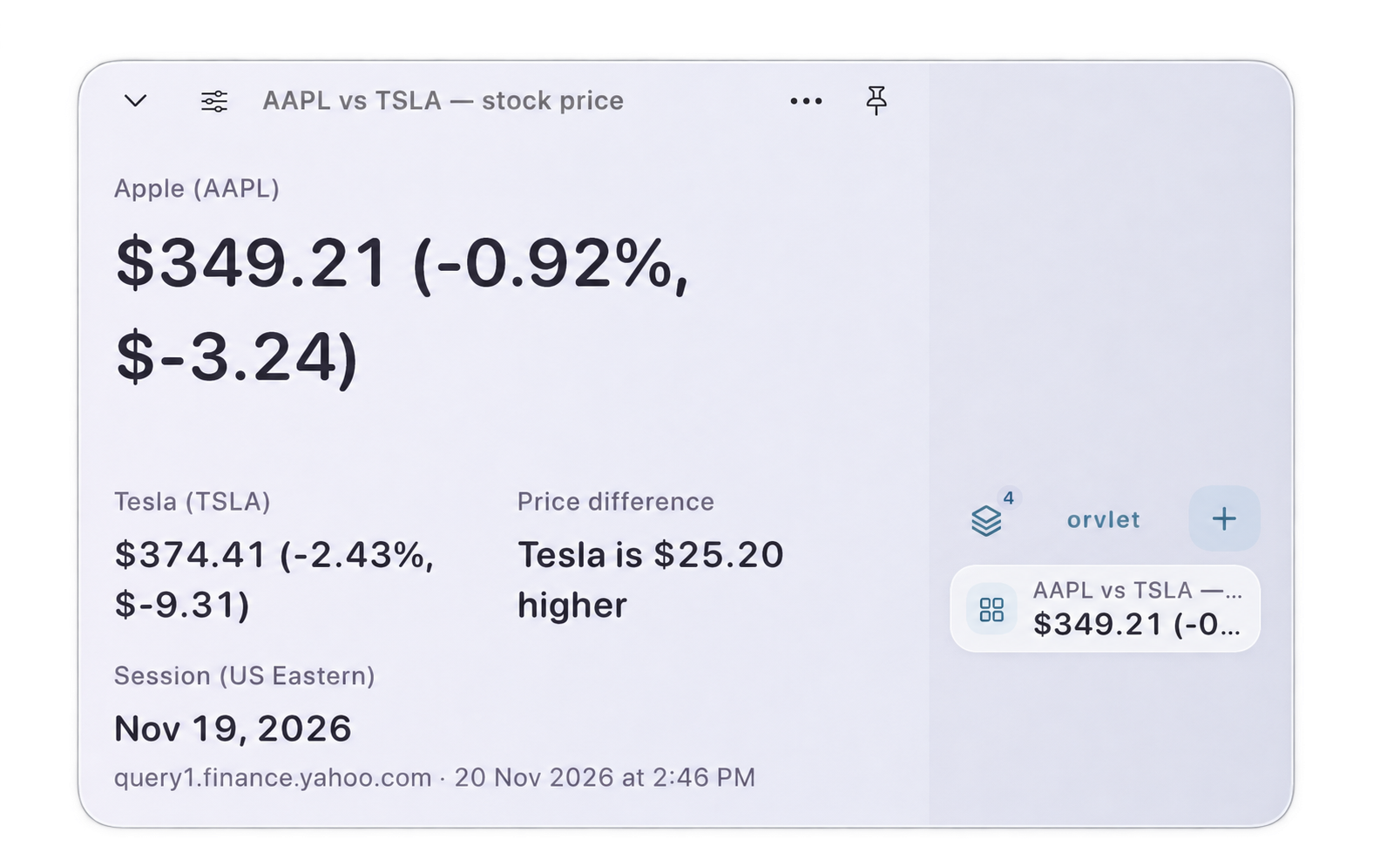} \\
        \small (c) GitHub Star Count &
        \small (d) Stock Price Comparison \\[6pt]

        \includegraphics[width=0.95\linewidth,height=1.5cm,keepaspectratio]{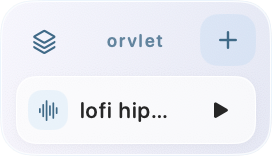} &
        \includegraphics[width=0.95\linewidth,height=3cm,keepaspectratio]{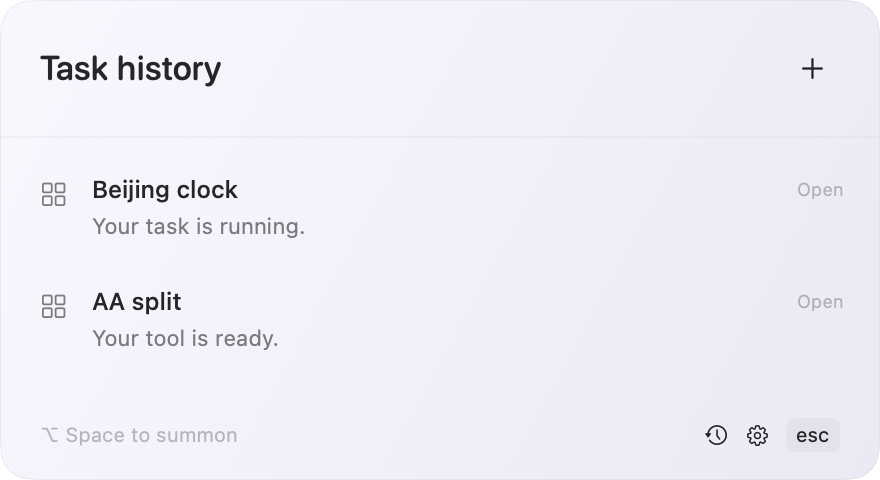} \\
        \small (e) Collapsed Edge Pill &
        \small (f) Task History
    \end{tabularx}
    \caption{Task interfaces in \sys. (a) Music card with playback and volume controls. (b) Dual-timezone clock card. (c) Card for checking a GitHub repository's star count. (d) Card comparing Apple and Tesla share prices. (e) Collapsed edge pill. (f) Task history view for returning to ongoing tasks.}
    \label{fig:surfaces}
\end{figure*}

Each active task appears as a small pill at the workspace edge, with a label, status, and main action (Fig.~\ref{fig:surfaces}e). When the interface is collapsed and neither interaction nor feedback requires attention, the pills fade to a semi-transparent appearance after three seconds. This reduces their visual prominence to limit distraction from primary work. They return to full opacity when the pointer enters or interaction resumes. Hovering over or opening a pill reveals a compact card. This provides different levels of access: a user can read a status, use a main action, or open additional controls according to the current need.

The system generates a combination of information and reusable controls for the task. For P2, a user can pause playback through an existing control or read a countdown from the task interface, without writing another request or waiting for a new interface. These actions use the current task's controls and displayed values. For P3, the playback status and remaining timer value provide concrete information when the user returns, alongside controls for continuing interaction. A task history view offers another entry point to existing tasks (Fig.~\ref{fig:surfaces}f).

\sys also supports clarification when essential user information is missing. The agent can ask a concise follow-up question and mark the request as needing input. The question and reply remain associated with the same task. When its conversation is not open, the question appears as feedback at the workspace edge without automatically focusing the input field. The feedback remains available until read or dismissed. This further supports P3 by showing what information the user needs to provide to continue the task.

% SEEV offers a qualitative lens for these interface choices \cite{steelman2017theory}. Idle fading lowers salience, while a persistent edge location is intended to reduce the effort of locating task information. The expectancy and value factors also suggest that checking depends on how often useful changes are anticipated and how much they matter to the current task. Compactness alone therefore does not determine the attention an interface requires.

\subsection{Scenarios}
\label{sec:walkthrough}

\paragraph{Controlling Music.}
A user asks to play quiet background music and returns to writing. \sys starts playback in a predefined music card with playback and volume controls (Fig.~\ref{fig:surfaces}a). 

\paragraph{Monitoring Concurrent Time-Related Tasks.}
A user coordinating work between Beijing and London asks for a dual-timezone clock and a break countdown. \sys presents the detailed timezone view in a card while stacking pills for the timezone summary and countdown alongside primary work (Fig.~\ref{fig:surfaces}b). 

\paragraph{Checking and Refreshing a Repository's Star Count.}
A user asks for a GitHub repository's star count. \sys presents the count, repository identifier, data source, and a refresh control in a compact card, paired with a summary pill (Fig.~\ref{fig:surfaces}c).

\paragraph{Comparing Stock Prices.}
A user asks to compare Apple and Tesla share prices. \sys presents both prices, their changes, and the price difference in a single card alongside the trading session, data source, and a summary pill (Fig.~\ref{fig:surfaces}d). 

\subsection{Interaction Implications}
Table~\ref{tab:comparison} contrasts representative interaction configurations for an ongoing auxiliary task. The comparison identifies where users access controls and recover task context. These configurations can be combined; for example, a conversational interface can also provide persistent task controls.

\begin{table*}[t]
\centering
\caption{Representative interaction configurations for ongoing auxiliary tasks. }
\label{tab:comparison}
\small
\begin{tabularx}{\textwidth}{@{} >{\raggedright\arraybackslash}p{0.15\textwidth} Y Y Y Y @{}}
\toprule
Interaction need & Application controls & Chat bot & Generated task workspace & Calm Agent \\
\midrule
Establish a task & Navigate to relevant functions & Describe a goal & Describe a goal and generate its interface & Describe a goal for background execution \\
\addlinespace[4pt]
Routine adjustment & Use application controls & Write a request and read the reply & Use generated controls & Use existing controls in a pill or card \\
\addlinespace[4pt]
Inspect state & Open the relevant application view & Read an update or request one & Inspect the task workspace & Read a task indicator or expand its card \\
\addlinespace[4pt]
Return after absence & Locate the relevant view and state & Recover context from messages or a summary & Revisit the generated workspace & Access the continuing task's state and controls \\
\bottomrule
\end{tabularx}
\end{table*}

The design question is how much work a brief visit requires: locating the task, understanding its state, choosing an action, and returning to primary work. \sys brings these activities into a persistent point of access. The challenges below concern extending this approach across changing tasks, execution conditions, and user habits.

\section{Challenges}
\label{sec:challenges}

Calm Agent places competing demands on a single interface: it must be compact yet informative, adaptable yet familiar, responsive during ongoing execution, and able to request input without repeatedly disrupting primary work. We identify four open challenges in meeting these requirements.

\subsection{Adaptive Interfaces with Interaction Continuity}
A compact interface offers limited space for open-ended tasks with many possible states and actions. Showing only frequent controls risks hiding rare but necessary decisions, whereas exposing every option increases the effort of a brief visit. While evolving task-driven interfaces provide a foundation for adaptation \cite{cao2025generative}, supporting intermittent return requires deciding which elements should adapt and which should remain stable.

As user intentions and execution environments evolve, the system must determine whether a revision alters an action's meaning, its target, or merely a parameter. It must avoid both visually unchanged controls whose effects have shifted and unnecessary redesigns that discard familiarity. Progressive disclosure can keep surfaces compact, yet critical state distinctions, such as a blocked task or an unavailable action, must remain immediately legible to users who did not observe the transition.

\subsection{Attention Requests across Concurrent Tasks}
\label{sec:attention}
\label{sec:concurrent}
An agent must decide both whether human input is necessary and how prominently to request it. Asking about every ambiguity shifts interpretation work back to the user, whereas acting on unsupported assumptions risks unwanted outcomes; distinguishing missing information that changes the agent's action from choices resolvable within the delegated scope remains open. Even when input is needed, its timing and prominence should reflect the cost of delay and the user's current activity \cite{iqbal2007disruption,parasuraman2000model}: deferrable questions can remain quietly accessible, while time-sensitive decisions warrant stronger signals.

Concurrency compounds this problem across tasks. Individually reasonable requests may arrive together or interrupt primary work in succession, a tension made concrete by the stacked pills in \sys. Grouping requests lowers interruption frequency but risks delaying urgent decisions, whereas prioritizing one task makes others easy to overlook. Coordinating timing and salience across concurrent tasks without obscuring unresolved decisions remains a key challenge.

\subsection{Responsive Interface Generation and Execution}
\label{sec:responsiveness}
A brief interaction turns into a long wait if interface generation begins only after planning and execution finish. Preparing the interface and its underlying behavior in parallel, such as drafting a card while the agent resolves a data source, can reduce latency, but these processes share dependencies: valid actions and values often remain unknown until execution produces evidence.

A staged interface can surface received intent and known information early, enabling controls as their backing behavior becomes ready. Doing so requires clearly distinguishing pending operations from ready controls and reconciling partial results whenever the user revises the goal mid-execution. Without careful coordination, asynchronous updates risk causing layout shifts or invalid actions just as the user attempts to interact.

\subsection{Understanding and Adapting to Users over Time}
\label{sec:personalization}
Repeated use enables an agent to interpret requests with less explanation and tailor interfaces to user habits, such as preferred terminology, frequent controls, or level of detail. Yet brief interactions provide incomplete and ambiguous evidence: a control may be used often simply because it is easy to reach, and an ignored notification does not necessarily indicate a preference for silence.

Long-term adaptation must therefore separate explicit preferences from inferred habits and account for task context. A preference for minimal displays during focused work may not apply when exploring a task in detail, and explicit corrections should revise the current assumption without erasing habits that remain useful elsewhere. Deciding which learned preferences transfer to a new request, while tracking their confidence and keeping assumptions easy to correct, remains an open problem.

\section{Conclusion}
\label{sec:conclusion}

This paper presents Calm Agent, a design framework for ongoing auxiliary tasks under intermittent human attention. Its contribution is to connect task continuity with three on-demand interaction modes: Delegate, Adjust, and Check. A persistent task representation links these modes to execution, while agent-initiated requests bring users back when clarification or a decision is needed. The Orvlet prototype illustrates this framework through background execution, compact task controls, and task-linked clarification. We identify four open challenges in interface adaptation and continuity, attention requests across concurrent tasks, responsive generation and execution, and long-term user adaptation. Together, these contributions frame agent interaction around brief, purposeful participation while protecting attention for primary work.

\bibliographystyle{unsrtnat}
\bibliography{lib}

\end{document}